# S-band single-longitudinal-mode erbium-doped fiber ring laser with ultra-narrow linewidth, ultra-high OSNR, high stability and low RIN

Zhengkang Wang[a], Jianming Shang[b], Siqiao Li[a], Kuanlin Mu[a], Yaojun Qiao [a,*] and Song Yu[b]

[a]*Beijing Key Laboratory of Space-Ground Interconnection and Convergence, School of Information and Communication Engineering, Beijing University of Posts and Telecommunications, Beijing 100876, China*
[b]*Institute of Information Photonics and Optical communications, Beijing University of Posts and Telecommunications, Beijing 100876, China*

**Abstract**

A high-performance S-band single-longitudinal-mode (SLM) erbium-doped fiber (EDF) ring cavity laser based on a depressed cladding EDF is investigated and experimentally demonstrated. We combine a double-ring passive resonator (DR-PR) and a length of unpumped polarization maintaining (PM) EDF in the laser cavity to achieve the SLM lasing without mode hopping. The DR-PR, composed of two efficient dual-coupler fiber rings, is utilized to expand the free spectral range of the EDF ring cavity laser and to eliminate the dense longitudinal modes greatly. The PM EDF, insusceptible to random change induced by environmental perturbations, is used as a saturable absorber filter to guarantee and to stabilize the SLM operation of the EDF ring cavity laser. At the pump power of 400 mW, we obtain an SLM EDF ring laser with a linewidth as narrow as 568 Hz, an optical signal-to-noise ratio as high as 77 dB, and a relative intensity noise as low as 140 dB/Hz at the frequency over 5 MHz. Meanwhile, the stability performance of both the wavelength lasing and the output power, the dependence of the OSNR and the output power on pump power for the S-band fiber laser are also investigated in detail.

k*eywords:* S-band, single longitudinal mode (SLM), erbium-doped fiber (EDF), fiber ring laser.

*Corresponding author
 *Email addresses:* qiao@bupt.edu.cn (Yaojun Qiao)

## 1. Introduction

The S-band erbium-doped fiber laser (EDFL), a laser resource as important as the C- and L-band EDFLs, has not been given enough research attention in contrast with the fact that the C- and L-band EDFLs have been extensively studied. At present, in the fields of the C- and L-band EDFLs, the single-longitudinal-mode (SLM) EDFLs have drawn abundant attention due to their potential value of research and applications. The configurations reported to achieve SLM operation of EDFLs mainly include DFB [1-2], DBR [3] and ring cavity [4–6]. Among them, SLM erbium-doped fiber (EDF) ring lasers with narrow linewidth, high optical signal to noise ratio (OSNR), easy operability, high stability and low relative intensity noise (RIN) have been recognized as the preferred light sources for fields such as optical communication, optical metrology, atomic spectroscopy and high-resolution spectroscopy [7-10], to enable better accuracy and higher stability. To deepen the study, many proposals with substantial techniques and optical components have been put forward to solve the problem of obtaining a stable SLM EDF ring laser, such as the compound-cavity structure [5-6], the unpumped EDF-based saturable absorber (SA) filter [11], the optical injection method [12], the Mazh-zehnder interferometer [13], the self- and external-injection technologies in the EDF [14-15], the ultra-narrow pass-band filter [16] and the phase shifted fiber Bragg grating (FBG) [17]. By contrast, we can find very few studies on the S-band EDFLs [18-20], let alone those on the S-band ring cavity lasers [18-19]. Even with the published research on the S-band ring cavity lasers, we can find some weaknesses, such as low OSNR, wide linewidth and insufficient stability.

Although the C- and L-band EDFLs, especially the ring cavity lasers, have been the frequent study focus, there are still some improvement to be made with their OSNR value, their linewidth and their stability performance. In our previous C-band study [4], for instance, the OSNR value of the EDFL is affected by the insufficient mode selecting capability of the passive subring resonator, which is caused by the low coupling ratios of the OCs and the short lengths of the passive subring resonator. In addition, the stability of the distributed gain grating formed in the unpumped single mode EDF is affected since the state of polarization in the EDF is susceptible to random change induced by environmental perturbation, which affects the linewidth and the stability performance of the EDFL.

In this paper, we attempt to undertake a study to obtain a S-band EDF ring cavity laser based on an erbium-doped silica fiber with depressed cladding design to meet the requirement of a stable S-band light source. The depressed cladding EDF is employed to extend the operating wavelengths of EDF ring cavity laser to S-band range thanks to the EDF gain extension effect. We combine a high-quality double-ring passive resonator (DR-PR) and a length of unpumped polarization maintaining (PM) EDF to improve the OSNR value, the linewidth the stability performance of the EDFL. The DR-PR, composed of two dual-coupler fiber rings, serves as a mode filter with excellent mode selecting capability to eliminate the dense longitudinal modes of the EDFL and to benefit the SLM operation. The PM EDF, functioning as an SA filter with an ultra-narrow bandwidth, is utilized to guarantee the SLM operation and to narrow the linewidth of the EDFL. In addition, the PM EDF can increase the stability and prolong the mode-hopping-free SLM operation time of the EDFL, because of its insensitivity to the environment disturbance. In short, with the application of the DR-PR and the PM EDF, we achieve an S-band SLM EDFL with an ultra-high OSNR, ultra-narrow linewidth, high stability and low RIN.

## 2. Experimental setup and principles

The experimental setup of the demonstrated S-band SLM fiber laser is illustrated in Fig. 1, which mainly consists of a 30-m-long depressed-cladding EDF (Er:Al/Ge/SiO2 fiber), a circulator (CIR), a fiber Bragg grating (FBG), a polarization beam splitter (PBS), a DR-PR and a 4.5-m-long PM EDF (IXF-EDF-FGL-1480-PM). In the experiment, the depressed-cladding EDF [18] serves as the gain medium pumped by a 980 nm laser diode (LD) through a 980/1500 nm wavelength division multiplexer (WDM). The FBG serves as an excellent mode restricting element to reduce the potential longitudinal mode in the EDFL. The center-reflecting wavelength, the reflectivity and the 3-dB bandwidth of the FBG are 1499.781 nm, 94.88% and 0.196 nm. The isolator (ISO) and the CIR are employed in the main cavity to guarantee the unidirectional operation of the proposed EDFL to avoid the MLM oscillation caused by the spatial-hole burning. The PBS is exploited to function as a mode filter to suppress the mode competition and to divide the light beam into two orthogonal linearly polarized light beams. The generated laser is extracted from the 40% port of a 60:40 optical coupler (OC3).

The DR-PR, which is inserted between the third port of CIR and the input port of PBS, is composed of two optical couplers (OC1 and OC2) with the coupling ratio of 90:10, is employed to serve as a high-quality mode filter to eliminate the dense longitudinal modes of the S-band EDFL. The 4.5-m unpumped PM EDF inserted between the second port of CIR and FBG, is utilized as an SA filter to guarantee and to stabilize the SLM operation of the S-band EDFL.

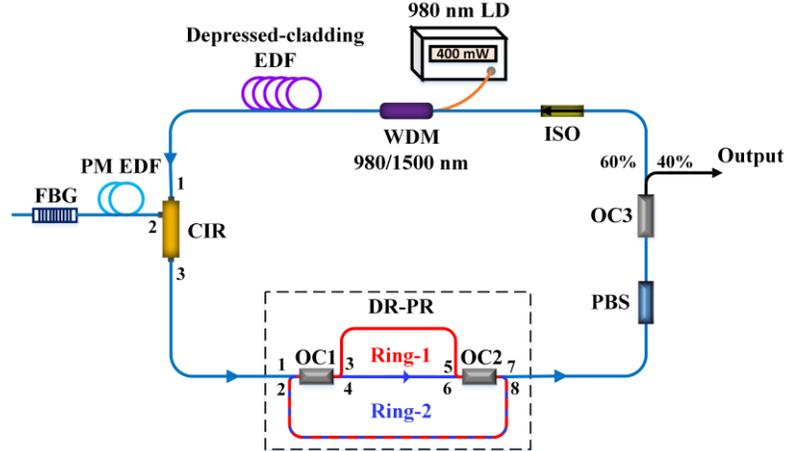

Fig. 1 Experimental setup of the designed S-band EDFL. OC, optical coupler; LD, laser diode; WDM, wavelength division multiplexer; EDF, erbium-doped fiber; FBG, fiber Bragg grating; CIR, circulator; ISO, isolator; PBS, polarization beam splitter; DR-PR, double-ring passive subring resonator.

The main cavity length of the proposed EDFL is ~39.5 m, corresponding to a longitudinal-mode spacing of 5.2 MHz. However, the 3-dB bandwidth of the FBG is 0.196 nm corresponding to a frequency bandwidth of 24.5 GHz at 1500 nm. In order to obtain the SLM operation, the SLM selecting element is required to guarantee that only one longitudinal mode to be selected from more than 4700 modes. In our previous work [4], we proved that the proposed multiple-ring compound resonator had excellent SLM selection capabilities. In this paper, we further improved the coupling ratios of the OCs and the lengths of the double rings (Ring-1 and Ring-2) to obtain an DR-PR with better SLM selecting performance.

The SLM selecting mechanism of the proposed DR-PR is shown as follows. As seen from the Fig. 1, when the input light beams are injected into OC1, 10% of the output light beams go straight into the fifth port of the OC2, and 90% of the output light beams go into the sixth port of the OC2. Then, among the output light beams that pass OC2, 90% feed back to OC1, while 10% serve as the output of the DR-PR. As can be seen in Fig. 2, OC1 and OC2 form a double-ring architecture (Ring-1 and Ring-2) with the lengths of 1.9 m and 2.6 m and the corresponding FSRs of Ring-1 and Ring-2 can be calculated as 107.3 MHz and 78.4 MHz according to the equation:

$$FSR = c/nL \qquad (1)$$

where $c$ is the light speed in vacuum, $n$ is the effective refractive index at 1500 nm of single-mode fiber and $L$ is the length of the single-longitudinal fiber ring. Then the effective FSR of the proposed DR-PR can be calculated to be ~8.4 GHz in accordance with the Vernier effect.

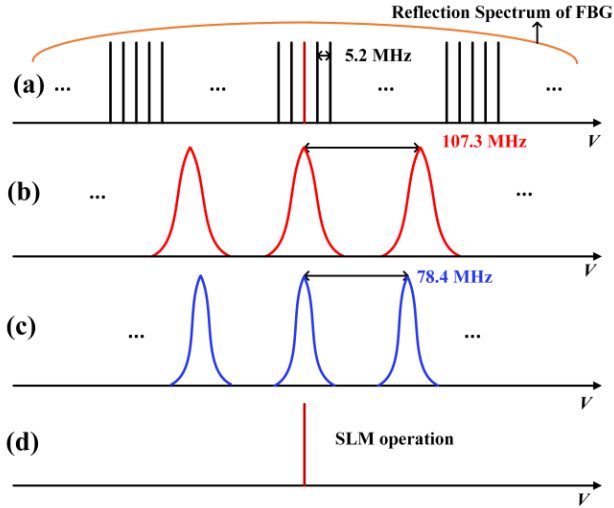

Fig. 2 Schematic diagram of the SLM operation in the S-band EDFL. (a) Dense longitudinal modes of the main ring cavity and the reflection spectrum of the FBG; (b) The FSR of the Ring-1; (c) The FSR of the Ring-2; (d) The selected SLM.

As shown in Fig. 2, the proposed DR-PR serves as a core SLM selecting element, and the number of longitudinal modes in the laser cavity can be decreased effectively due to the Vernier effect. Compared with our previous work [4], the coupling ratios of the OCs and the lengths of the double rings (Ring-1 and Ring-2) are further improved to benefit the proposed DR-PR function as a more effective mode filter to suppress the dense MLM oscillation of the EDFL. However, the effective FSR of the DR-PR is much smaller than the 3-dB reflecting bandwidth of the FBG and there are more than one pass band located in the 3-dB bandwidth of the FBG. Thus, the proposed DR-PR cannot guarantee the SLM operation of the EDFL.

To achieve the stable SLM operation of the S-band EDFL, a length of unpumped PM EDF is utilized in the laser cavity to guarantee and to stabilize the SLM operation.

As shown in Fig. 1, two counter-propagating light beams pass through the PM EDF and a distributed gain grating with an ultra-narrow bandwidth can form in the PM EDF by interfering between two counter-propagating waves. The distributed gain grating serves as a self-tracking filter, and the full width at half maximum (FWHM) of the self-tracking filter is calculated to be ~4.1 MHz according to the following equation [21]:

$$\Delta f = \frac{c}{\lambda} \frac{2\Delta n}{n_{eff} \lambda} \sqrt{(\frac{\Delta n}{2 n_{eff}})^2 + (\frac{\lambda}{2 n_{eff} L_g})^2} \qquad (2)$$

where $\Delta n$ is the variation of the refractive index, which can be given by the Kramers-Kronig relation [22]. $n_{eff}$ is the effective index of the unpumped PM EDF, $\lambda$ is the central wavelength and $L_g$ is the length of the unpumped EDF2 in the DR-PR.

## 3. Experimental results

For the S-band EDFL system which was just built and constructed on an ordinary optical table, we carried out all experiments at laboratory temperature and without extra vibration isolation and temperature compensation techniques. The S-band EDFL presented an ultra-high OSNR, ultra-narrow linewidth, high stability and low RIN, which are demonstrated as follows.

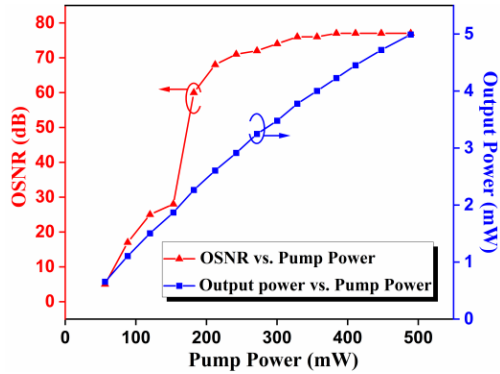

Fig. 3. The OSNR value and Output power variations versus pump power

Firstly, the relationships between the OSNR value and output power variations versus the pump power were measured by a power meter (ILX Lightwave) with the resolution of 0.001 dB and an optical spectrum analyzer (OSA, AQ637OD) with the resolution of 0.02 nm, as shown in Fig. 3. The lasing threshold of the S-band EDFL was about 56 mW with the OSNR of ~5 dB. As can be seen, the laser output power was linearly enhanced with the pump power increasing from 56 mW to ~500 mW and the saturation of the output power was not observed. However, when the pump power was bigger than 300 mW, the OSNRs of the S-band EDFL had an increasing trend up to ~77 dB. To obtain the stable SLM lasing of the EDFL and to protect the 980 LD, the optical spectrum, the SLM characteristic, the linewidth and the stability performance of the S-band EDFL were investigated when the pump power was fixed at 400 mW, which are demonstrated as follows.

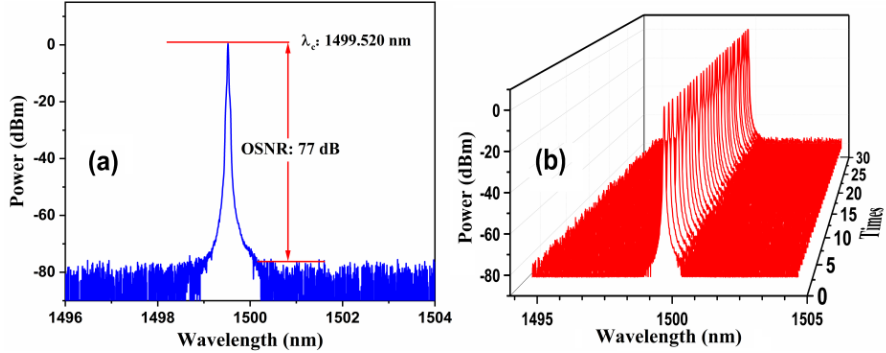

Fig. 4 The optical spectrum measurements: (a) optical spectrum of the laser in SLM operation and (b) spectra of 30 times repeated OSA scans at 1-min intervals.

Keeping the pump power at 400 mW, the optical spectrum of the EDFL was measured by the OSA with the resolution of 0.02 nm as shown in Fig. 4(a). The center wavelength of the EDFL was 1499.520 nm with an ultra-high OSNR value of 77 dB. The ultra-high OSNR value indicates a high oscillating quality of the S-band EDFL because of the outstanding mode selection capability of the designed DR-PR. In addition, as shown in Fig. 4(b), the three-dimensional plot shows the variations of wavelength and output power of the EDFL taken from the spectra of first 30-min observation time. Since there were little or no fluctuations of either the output power or wavelength, displaying a stable SLM operation of the proposed EDFL.

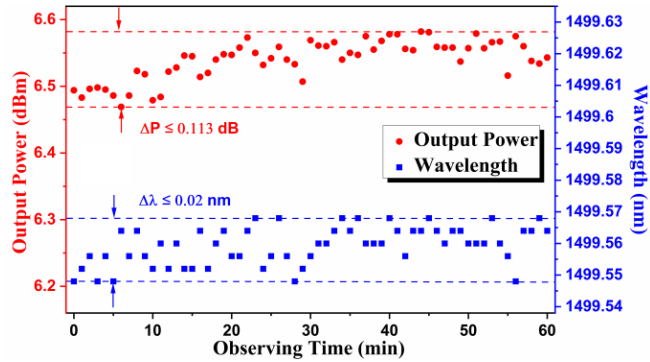

Fig. 5 The stability performances of the output power and the lasing wavelength at pump power of 400 mW in a 1-min interval over 1 h.

To further investigate the stability performance of the EDFL, the variations of the center wavelength and the output power were measured through an OSA and a power meter during a 1-hour experiment with the interval of 1-min, as shown in Fig. 5. The lasing wavelength variation and the output power fluctuation of the S-band EDFL were less than 0.02 nm and 0.013 dB, suggesting a very stable operation of the EDFL mainly based on the excellent mode-selection capability of the DR-PR and the insusceptibility of the PM SA filter to random change induced by environmental perturbations.

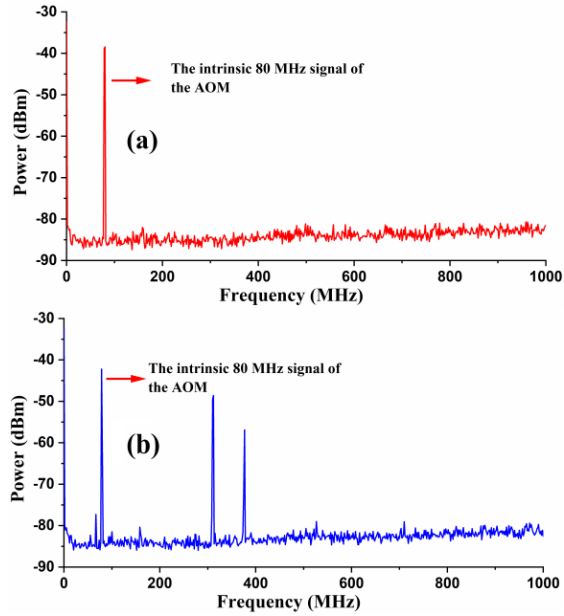

Fig.6. The radio frequency beating spectra measured by delayed self-heterodyne measurement system in: (a) 0~1 GHz span with 100-kHz resolution bandwidth; (b) 0~1 GHz span without unpumped PM-EDF with 100-kHz resolution bandwidth;

The SLM operation of the EDFL was verified through the delayed self-heterodyne system, which is composed of a 1-GHz photodetector (PD), a RF electrical spectrum analyzer (ESA) (Rohde & Schwarz FSV), a Mach-Zehnder interferometer with an 80 MHz acoustic optical modulator (AOM) and a 47-km-long single mode fiber delay line in two arms respectively. As shown in Fig. 6(a), there are no other obvious beating signals captured except for the intrinsic 80 MHz sifted signal of the AOM in the 1-GHz range, showing the proposed EDFL was in SLM operation. Meanwhile, the SLM operation could operate stably in 2-h monitoring time, indicating the EDFL was in stable SLM operation. To investigate the SLM selecting capability of the PM EDF-based SA filter, we removed the PM EDF and MLM oscillation was captured as shown in Fig. 6(b), which justifies the excellent SLM selecting capability of the SA filter.

Then, the spectral linewidth of the EDFL was also measured through the delayed self-heterodyne system [23]. Fig. 7 shows the EDFL linewidth with 300-Hz resolution bandwidth in 79.85~80.15 MHz. To measure the spectral linewidth precisely, the measured data was fitted with Lorentzian function. As can be seen, the measured date fitted well with the Lorentzian lineshape. In addition, the 20-dB bandwidth of ~11.36 kHz was obtained, meaning that the real linewidth of the EDFL is as narrow as 568 Hz since the linewidth of the proposed EDFL is ~1/20 of the 20-dB bandwidth of the corresponding fit curve. The excellent linewidth performance of the EDFL mainly benefits from the extremely long main cavity length of the EDFL (~40 m) and the linewidth narrowing capability of the SA filter. It should be noticed that it is impossible to obtain a pure Lorentzian linewidth spectrum (using a delay fiber over 1000 km) due to the serious 1/f frequency noise and the limitation of output power of the EDFL. Thus, the estimated linewidth of 568 Hz could be regarded as conservative characterization of the laser linewidth.

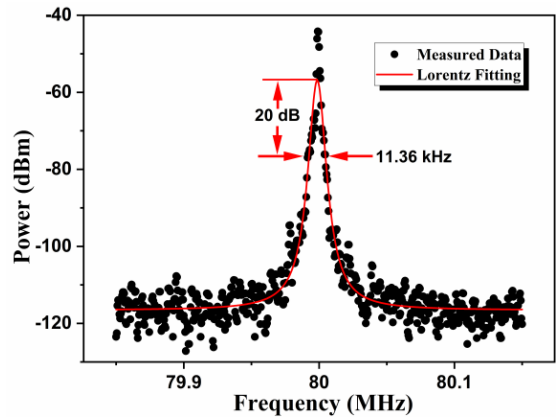

Fig. 7 The linewidth measurement of the EDFL with 300 Hz resolution bandwidth.

Lastly, the RIN characteristic of the S-band EDFL was measured and investigated. As shown in Fig. 8, the RIN decreased from −107 dB/Hz to −140 dB/Hz while frequency increased and the RIN was stabilized at approximately −140 dB/Hz for frequencies over 5 MHz, enabling the proposed EDFL for many high precision applications. It is worth noticing that the relaxation oscillation frequency peak of −107 dB/Hz-quality was observed at around 110 kHz for the proposed EDF, which was mainly caused by the external disturbances or the measurement system. However, the relaxation oscillation frequency of the EDFL wasn't observed at the frequencies of >5

MHz because of the outstanding mode-selection capability of the designed DR-PR and the excellent capability of the PM EDF to stabilize the SLM oscillation of the EDFL.

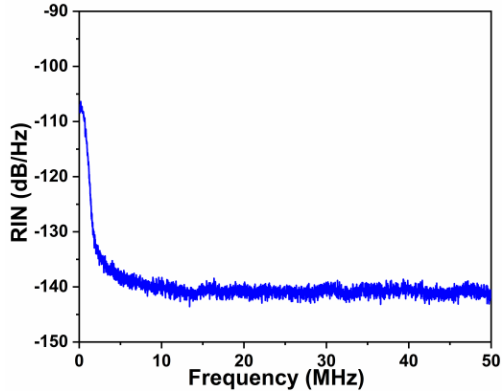

Fig. 8 The RIN spectrum of the EDFL.

## 4. Conclusions

In summary, we have presented and designed an S-band SLM EDFL with an ultra-narrow linewidth, an ultra-high OSNR, high stability and low RIN based on a depressed cladding EDF. The depressed cladding EDF is utilized to extend the operating wavelength of the EDFL to S-band range. A high-quality DR-PR and a PM EDF-based SA filter are collectively utilized in the proposed EDFL to suppress the dense MLM oscillation and to guarantee the SLM operation. Compared with our previous study [4], the designed DR-PR presents higher mode-selection capability since the coupling ratios and the cavity lengths of two dual-couple fiber rings formed in the DR-PR have been further improved. In addition, the PM EDF increases the stability of the EDFL thanks to its insusceptibility to environment disturbance. In the SLM operating mode, at the pump power of 400 mW, the laser linewidth is as narrow as 568 Hz, the OSNR value is as high as ~77 dB and the RIN is as low as -140 dB/Hz for frequencies above 5 MHz. This kind of EDFL can serve as a stable S-band light source with excellent performance to meet the requirement for a stable S-band light source in many important applications.

**Acknowledgements**

This paper is supported by Natural Science Foundation of China (NSFC) (61690195 and 61701040); Beijing University of Posts and Telecommunications (BUPT) Action Project for Promoting the Development of Scientific and

**References**

[1] L. Dong, W. H. Loh, J. E. Caplen, J. D. Minelly, K. Hsu, and L. Reekie, "Efficient single-frequency fiber lasers with novel photosensitive Er/Yb optical fibers," Opt. Lett., vol. 22, no. 10, pp. 694–696, 1997.

[2] Q. Li, F. Yan, W. Peng, T. Feng, S. Feng, S. Tan, P. Liu, and W. Ren, "DFB laser based on single mode large effective area heavy concentration EDF," Opt. Express., vol. 20, no. 21, pp. 23684–23689, 2012.

[3] S. Xu, Z. Yang, W. Zhang, X. Wei, Q. Qian, D. Chen, Q. Zhang, S. Shen, M. Peng, and J. Qiu, "400 mW ultrashort cavity low-noise single-frequency $Yb^{3+}$-doped phosphate fiber laser," Opt. Lett., vol. 36, no.18, pp. 3708–3710, 2011.

[4] Z. K. Wang, J. M. Shang, L. H. Tang, K. L. Mu, S. Yu, and Y. J. Qiao, "Stable Single-Longitudinal-Mode Fiber Laser With Ultra-Narrow Linewidth Based on Convex-Shaped Fiber Ring and Sagnac Loop," IEEE Access., vol. 7, no. 1, pp. 166398–166403, 2019.

[5] S. Feng, Q. Mao, Y. Tian, Y. Ma, W. Li, and L. Wei, "Widely tunable single longitudinal mode fiber laser with cascaded fiber-ring secondary cavity," IEEE Photonics Technol. Lett., vol. 25, no. 4, pp. 323–326, 2013.

[6] Z. K. Wang, J. M. Shang, K. L. Mu, S. Yu and Y. J. Qiao, "Single-longitudinal-mode fiber laser with an ultra-narrow linewidth and extremely high stability obtained by utilizing a triple-ring passive subring resonator," Optics & Laser Technology., vol. 130, Art. no. 106329, 2020.

[7] S. Diaz, D. Leandro, and M. Lopez-Amo, "Stable multiwavelength erbium fiber ring laser with optical feedback for remote sensing," J. Lightwave Technol., vol. 33, no. 12, pp. 2439–2444, 2015.

[8] C. S. Kim, F. Farokhrooz, and J. Kang, "Electro-optic wavelength-tunable fiber ring laser based on cascaded composite sagnac loop filters," Opt. Lett., vol. 29, no. 14, pp. 1677–1679, 2004.

[9] T. Wu, X. Peng, W. Gong, Y. Zhan, Z. S. Lin, B. Luo, and H. Guo, "Observation and optimization of 4He atomic polarization spectroscopy," Opt. Lett., vol. 38, no. 6, pp. 986–988, 2013.

[10] B. Liu, C. Jia, H. Zhang, and J. Luo, "DBR-fiber-laser-based active temperature sensor and its applications in the measurement of fiber birefringence," Microw. Opt. Technol. Lett., vol. 52, no. 1, pp. 41–44, 2010.

[11] X. He, D. N. Wang, and C. R. Liao, "Tunable and switchable dual-wavelength single-longitudinal-mode erbium doped fiber lasers," J. Lightwave Technol., vol. 29, no. 6, pp. 842–849, 2011.

[12] C. H. Yeh, T. J. Huang, Z. Q. Yang, C. W. Chow, and J. H. Chen, "Stable single-longitudinal-mode erbium fiber ring laser utilizing self-injection and saturable absorber," IEEE Photonics J., vol. 9, no. 6, Art. no. 7106206, 2017.

[13] M. I. MdAli, S. A. Ibrahim, M. H. Abu Bakar, A. S. M. Noor, S. B. Ahmad Anas, A. K. Zamzuri, and M. A. Mahdi, "Tapered-EDF-based Mach Zehnder interferometer for dual-wavelength fiber laser," IEEE Photon. J., vol. 6, no. 5, pp. 1-9, 2014.

[14] X. Zhang, N. H. Zhu, L. Xie, B. X. Feng, "A stabilized and tunable single-frequency erbium-doped fiber ring laser employing external injection locking," J. Lightwave Technol., vol. 25, no. 4, pp.1027–1033, 2007.

[15] T. Zhu, L. Shi, S. Huang, "Ultra-narrow linewidth fiber laser with self-injection feedback based on Rayleigh backscattering, Proc. CLEO. 2, pp.1-2, 2014.

[16] H. Zou, S. Lou, G. Yin, and W. Su, "Switchable dual-wavelength PM-EDF ring laser based on a novel filter", IEEE Photon. Technol. Lett. vol. 25, no. 11, pp.1003–1006, 2013.

[17] X. Chen, J. Yao, F. Zeng, and Z. Deng, "Single-longitudinal-mode fiber ring laser employing an equivalent phase-shifted fiber Bragg grating," IEEE Photonics Technol. Lett., vol. 17, no. 7, pp. 1390–1392, 2005.

[18] C. H. Yeh, C. C. Lee, and S. Chi, "A Tunable S-Band Erbium-Doped Fiber Ring Laser," IEEE Photonics Technol. Lett., vol. 15, pp. 1053–1054, 2003.

[19] H. C. Chien, C. H. Yeh, C. C. Lee, and S. Chi, "A tunable and single-frequency s-band erbium fiber laser with saturable-absorber-based autotracking filter," Opt. Commun., vol. 250, pp. 163-167, 2005.

[20] S. H. Xu, Z. M. Yang, T. Liu, W. N. Zhang, Z. M. Feng, and Q. Y. Zhang, "An efficient compact 300 mw narrow-linewidth single frequency fiber laser at 1.5 microm," Opt. Exp., vol. 18, no. 2, pp. 1249-1254, 2010.

[21] T. Wang, L. Zhang, C. Feng, M. Qin, and L. Zhan, "Tunable bistability in hybrid Brillouin–erbium single-frequency fiber laser with saturable absorber", J. Opt. Soc. Amer. B., vol. 33, no. 8, pp. 1635–1639, 2016.

[22] S. Pan, and J. Yao, "Frequency-switchable microwave generation based on a dual-wavelength single-longitudinal-mode fiber laser incorporating a high-finesse ring filter," Opt. Exp., vol. 17, no. 14, pp. 12167–12173, 2009.

[23] H. L, M. T, and M. K, "Laser linewidth measurements using self-homodyne detection with short delay," Opt. Commun., vol. 155, pp. 180–186, 1998.